%% file: proceedings.tex
\documentclass[a4paper]{PoS}

\include{commands}
\usepackage{braket}
\usepackage{siunitx}
\usepackage{nicefrac}

\newcommand{\projector}[1]{\Gamma_{\!#1}}
\newcommand{\proj}[1]{\projector{\vect{#1}}}

\DeclareGraphicsExtensions{.eps,.pdf,.jpg,.png}
\graphicspath{{./graphs/}}

\title{Electromagnetic Form Factors of Nucleon Excitations in Lattice QCD}

\ShortTitle{Electromagnetic Form Factors of Nucleon Excitations}

\author{\speaker{Finn M. Stokes}, Waseem Kamleh, Derek B. Leinweber and Benjamin J. Owen\\
        Centre for the Subatomic Structure of Matter,\\
	Department of Physics,\\
        University of Adelaide, SA 5005\\
	E-mail: \email{finn.stokes@adelaide.edu.au}}

\abstract{Variational analysis techniques in lattice QCD are powerful tools that give access
to the full spectrum of QCD\@. At zero momentum, these techniques are well established and
can cleanly isolate energy eigenstates of either positive or negative parity.
In order to compute the form factors of a single energy eigenstate, we must perform a
variational analysis at non-zero momentum. When we do this with baryons, we run into issues
with parity mixing in the Dirac spinors, as boosted baryons are not eigenstates of parity.  Due to this parity mixing,
care must be taken to ensure that the projected correlation functions provided by the variational
analysis correspond to the same states at zero momentum. This can be achieved through the parity-expanded
variational analysis (PEVA) technique, a novel method developed at the University of Adelaide for ensuring
the successful and consistent isolation of boosted baryons.
Utilising this technique, we are able to compute the form factors of baryon excitations without
contamination from other states. I present world-first calculations of excited state nucleon form
factors using this new technique.}

\FullConference{The 26th International Nuclear Physics Conference\\
                11-16 September, 2016\\
                Adelaide, Australia}

\begin{document}

\section{Introduction}
To evaluate the form factors and transition moments of baryon excitations in lattice
QCD, it is necessary to isolate these states at finite momentum. Excited
baryons have been isolated on the lattice through a combination of parity projection and
variational analysis techniques \cite{Kiratidis:2015vpa, Mahbub:2013ala, Mahbub:2012ri, Mahbub:2010rm,
Edwards:2012fx, Edwards:2011jj, Lang:2016hnn, Lang:2012db, Alexandrou:2014mka}.
At zero momentum, these techniques are well established. However, at non-zero momentum, these
techniques admit opposite parity contaminations.

To resolve this issue, the Parity-Expanded Variational Analysis (PEVA) technique \cite{Menadue:2013kfi}
was developed. By introducing a novel Dirac projector and expanding the operator basis used to
construct the correlation matrix, we are able to isolate states of both parities at
finite momentum.

Utilising the PEVA technique, we are able to present here the world's first lattice QCD calculations of
nucleon excited state form factors free from opposite parity contaminations. Specifically,
the Sachs electromagnetic form factors of a localised negative parity nucleon
excitation are examined.
Furthermore, we clearly demonstrate the efficacy of variational analysis techniques at providing access
to ground state form factors with extremely good control over excited state effects.

\section{Parity-Expanded Variational Analysis}
\label{sec:PEVA}
The PEVA technique \cite{Menadue:2013kfi} was developed to solve the problem of opposite parity
contaminations at finite momentum. In this section, we briefly describe how it can be used
in calculations of baryon form factors.

The PEVA technique works by expanding the operator basis of the correlation matrix to isolate energy
eigenstates of both rest-frame parities simultaneously while still retaining a signature of this parity.
By considering the Dirac structure of the unprojected correlation matrix, we construct the
momentum-dependent projector \(\proj{\pm p} \definedby \frac{1}{4} \left(\identity + \gamma_4\right) \left(\identity \mp
\ii \gamma_5 \gamma_k \unitvect{p}_k\right)\). This allows us to construct a set of ``parity-signature''
projected operators \(\left\{\chi^i_{\pm\vect{p}} = \proj{\pm p} \, \chi^i \, , \; \chi^{i'}_{\pm \vect{p}} =
\proj{\pm p} \, \gamma_5 \, \chi^i\right\}\). The primed indices used here denote the inclusion of
\(\gamma_5\), inverting the way the operators transform under parity.

By performing a variational analysis with this expanded basis \cite{Menadue:2013kfi}, we construct optimised operators \(\phi^{\alpha}_{\pm\vect{p}}(x)\)
that couple to each state \(\alpha\). We can then use these operators to calculate the three point
correlation function
\begin{align*}
	\mathcal{G}^{\mu}_{\pm}(\vect{p'}, \vect{p};\, t_2, t_1;\, \alpha) \definedby &\sum_{\vect{x_2},\vect{x_1}} e^{-\ii \vect{p'}\cdot\vect{x_2}} \, e^{\ii (\vect{p'} - \vect{p})\cdot\vect{x_1}} \braket{\,\phi^{\alpha}_{\pm\vect{p'}}(x_2)\,|\,J^{\mu}(x_1)\,|\,\adjoint\phi^{\alpha}_{\pm\vect{p}}(0)\,}\,,
\end{align*}
where \(J^{\mu}\) is the \(O(a)\)-improved \cite{Martinelli:1990ny} conserved vector current used in
Ref.~\cite{Boinepalli:2006xd}, inserted with some three-momentum transfer \(\vect{q} = \vect{p'} - \vect{p}\).
We can take the spinor trace of this with some projector \(\Gamma\) to get the projected three point correlation function
\(G^{\mu}_{\pm}(\vect{p'}, \vect{p};\, t_2, t_1;\, \Gamma;\, \alpha) \definedby \tr\left(\Gamma \, \mathcal{G}^{\mu}_{\pm}(\vect{p'}, \vect{p};\, t_2, t_1;\, \alpha) \right)\).

We can then construct the reduced ratio,
\begin{align*}
	\adjoint{R}_{\pm}(\vect{p'}, \vect{p};\, \alpha;\, r, s) \definedby \, & \sqrt{\left|\frac{r_{\mu} \, G^{\mu}_{\pm}(\vect{p'}, \vect{p};\, t_2, t_1;\, s_{\nu} \, \projector{\nu};\, \alpha) \; r_{\rho} \, G^{\rho}_{\pm}(\vect{p}, \vect{p'};\, t_2, t_1;\, s_{\sigma} \, \projector{\sigma};\, \alpha)}{G(\vect{p'};\, t_2;\, \alpha) \, G(\vect{p};\, t_2;\, \alpha)}\right|} \\
	& \quad \times \mathrm{sign}\left(r_{\gamma} \, G^{\gamma}_{\pm}(\vect{p'}, \vect{p};\, t_2, t_1;\, s_{\delta}\, \Gamma_{\delta};\, \alpha)\right) \sqrt{\frac{2 E_{\alpha}(\vect{p})}{E_{\alpha}(\vect{p})+m_{\alpha}}} \, \sqrt{\frac{2 E_{\alpha}(\vect{p'})}{E_{\alpha}(\vect{p'})+m_{\alpha}}}\,,
\end{align*}
where \(r_{\mu}\) and \(s_{\mu}\) are coefficients selected to determine the form factors. By
investigating the \(r_{\mu}\) and \(s_{\mu}\) dependence of \(\adjoint{R}_{\pm}\), we find that the
clearest signals are given by
\begin{align*}
	R^{T}_{\pm} &= \frac{2}{1 \pm\, \unitvect{p} \cdot \unitvect{p'}} \; \adjoint{R}_{\pm}\left(\vect{p'}, \vect{p};\, \alpha;\, (1, \vect{0}), (1, \vect{0})\right) \,,\;\mathrm{and} \\
	R^{S}_{\mp} &= \frac{2}{1 \pm\, \unitvect{p} \cdot \unitvect{p'}} \; \adjoint{R}_{\mp}\left(\vect{p'}, \vect{p};\, \alpha;\, (0, \unitvect{r}), (0, \unitvect{s})\right) \,,
\end{align*}
where \(\unitvect{s}\) is chosen such that \(\vect{p} \cdot \unitvect{s} = 0 = \vect{p'} \cdot \unitvect{s}\),
\(\unitvect{r}\) is equal to \(\unitvect{q} \times \unitvect{s}\), and the sign \(\pm\) is chosen
such that \(1 \pm \unitvect{p} \cdot \unitvect{p'}\) is maximised.
We can then find the Sachs electric and magnetic form factors
\begin{align*}
	G_E(Q^2) =\, &\left[Q^2 \left(E_{\alpha}(\vect{p'}) + E_{\alpha}(\vect{p})\right)\, \left(\left(E_{\alpha}(\vect{p}) + m_{\alpha}\right)\,\left(E_{\alpha}(\vect{p'}) + m_{\alpha}\right) \mp \bigl|\vect{p}\bigr| \bigl|\vect{p'}\bigr|\right) \, R^{T}_\pm \right. \\
	& \left. \quad \pm\, 2 \bigl|\vect{q}\bigr| \left(1 \mp\, \unitvect{p} \cdot \unitvect{p'}\right) \bigl|\vect{p}\bigr| \bigl|\vect{p'}\bigr| \left(\left(E_{\alpha}(\vect{p}) + m_{\alpha}\right)\,\left(E_{\alpha}(\vect{p'}) + m_{\alpha}\right) \pm \bigl|\vect{p}\bigr| \bigl|\vect{p'}\bigr|\right) \, R^{S}_{\mp} \right] \\
	&/ \left[ 4 m_{\alpha} \left[ \left( E_{\alpha}(\vect{p}) \, E_{\alpha}(\vect{p'}) + m_{\alpha}^2 \mp\, \bigl|\vect{p}\bigr| \bigl|\vect{p'}\bigr| \right) \bigl|\vect{q}\bigr|^2 + 4 \bigl|\vect{p}\bigr|^2 \bigl|\vect{p'}\bigr|^2 \left(1 \mp\, \unitvect{p} \cdot \unitvect{p'}\right) \right] \right] \,,\;\mathrm{and} \\
	G_M(Q^2) =\, &\left[\pm\, 2 \left(1 \mp\, \unitvect{p} \cdot \unitvect{p'}\right) \bigl|\vect{p}\bigr| \bigl|\vect{p'}\bigr| \left(\left(E_{\alpha}(\vect{p}) + m_{\alpha}\right)\,\left(E_{\alpha}(\vect{p'}) + m_{\alpha}\right) \pm \bigl|\vect{p}\bigr| \bigl|\vect{p'}\bigr|\right) \, R^{T}_\pm \right. \\
	& \left. \quad - \bigl|\vect{q}\bigr| \left(E_{\alpha}(\vect{p'}) + E_{\alpha}(\vect{p})\right)\, \left(\left(E_{\alpha}(\vect{p}) + m_{\alpha}\right)\,\left(E_{\alpha}(\vect{p'}) + m_{\alpha}\right) \mp \bigl|\vect{p}\bigr| \bigl|\vect{p'}\bigr|\right) \, R^{S}_{\mp} \right] \\
	&/ \left[ 2 \left[ \left( E_{\alpha}(\vect{p}) \, E_{\alpha}(\vect{p'}) + m_{\alpha}^2 \mp\, \bigl|\vect{p}\bigr| \bigl|\vect{p'}\bigr| \right) \bigl|\vect{q}\bigr|^2 + 4 \bigl|\vect{p}\bigr|^2 \bigl|\vect{p'}\bigr|^2 \left(1 \mp\, \unitvect{p} \cdot \unitvect{p'}\right) \right] \right]\,.
\end{align*}
The details of this procedure will be presented in full in Ref.~\cite{stokes:formfactors}.

\section{Results}

Lattice QCD calculations of the Sachs electric and magnetic form factors of the
ground state nucleon and first negative-parity excitation are presented with good control over opposite
parity contaminations. The three-momentum of the current insertion is fixed. However, by varying the final
state momenta, we are able to gain access to these form factors
at a range of \(Q^2\) below \(|q_{min}|^2 = \SI{0.166(5)}{\giga\electronvolt\squared}\) and compare the \(Q^2\)
dependence to a dipole fit. The final state momenta considered and the corresponding \(Q^2\) values for both
states used here are presented in Table~\ref{tab:momenta}.

\begin{table}[b]
\caption{The final state and initial state momenta considered in units of the lattice
  momentum \(\frac{2\pi}{L_S}\) where \(L_S\)
  is the spatial length of the lattice, and the corresponding \(Q^2\) for each state in physical units.
  \label{tab:momenta}}
\begin{center}
\begin{tabular}{ccll}
&& Ground state & First -ve parity excitation \\
\(\vect{p'}\) (l.u.) & \(\vect{p}\) (l.u.) & \(Q^2\) (\si{\giga\electronvolt\squared}) & \(Q^2\) (\si{\giga\electronvolt\squared}) \\
\hline
\((1,0,0)\) & \((0,0,0)\) & 0.16028(14) & 0.16352(13) \\
\((1,0,1)\) & \((0,0,1)\) & 0.16097(11) & 0.16366(12) \\
\((2,0,0)\) & \((1,0,0)\) & 0.1255(8) & 0.1462(10) \\
\((2,0,1)\) & \((1,0,1)\) & 0.1295(7) & 0.1472(9) \\
\((3,0,0)\) & \((2,0,0)\) & 0.0872(11) & 0.1205(18) \\
\((3,0,1)\) & \((2,0,1)\) & 0.0928(10) & 0.1224(17) \\
\end{tabular}
\end{center}
\end{table}

These results are calculated on the second
lightest PACS-CS \((2+1)\)-flavour full-QCD ensemble \cite{Aoki:2008sm}, made available through the ILDG
\cite{Beckett:2009cb}. This ensemble uses a \(32^3 \times 64\) lattice, and employs an
Iwasaki gauge action with \(\beta = 1.90\) and non-perturbatively \(O(a)\)-improved Wilson quarks. We use the
\(m_{\pi}=\SI{296}{\mega\electronvolt}\) PACS-CS ensemble, and set the scale using the Sommer parameter with a physical value of
\(r_0 = \SI{0.4921(64)}{\femto\meter}\), giving a lattice spacing of \(a = \SI{0.0951(13)}{\femto\meter}\). With this scale,
our pion mass is \SI{280(4)}{\mega\electronvolt}. We used 367 gauge field configurations,
with two source locations on each configuration. The \(\chi^2 / \mathrm{dof}\) is calculated with the
full covariance matrix, and all fits have \(\chi^2 / \mathrm{dof}\ < 1.2\).

For the analyses in this section, we start with a basis of eight operators, by taking two conventional
spin-\(\nicefrac{1}{2}\) nucleon operators (\(\,\chi_1 = \epsilon^{abc} \, [{u^a}\transpose (C\gamma_5) \, d^b] \, u^c\),
and \(\chi_2 = \epsilon^{abc} \, [{u^a}\transpose (C) \, d^b] \, \gamma_5 \, u^c\,\)),
and applying 16, 35, 100, and 200 sweeps of gauge invariant Gaussian smearing when creating the propagators \cite{Mahbub:2013ala}.
For the conventional variational analysis, we take this basis of eight operators and project with \(\projector{\pm}\),
and for the PEVA analysis, we parity expand the basis to sixteen operators and project with \(\proj{p}\) and \(\proj{p}'\).

In extracting the form factors, we consider a fixed boundary condition in the time direction at \(t=N_t\). We fix the source at
time slice 16, and utilising sequential source techniques \cite{Bernard:1985ss} invert through the current, centring the
conserved current insertion at time slice 21 \cite{Leinweber:1990wn}. We choose time slice 21 by inspecting the two
point correlation functions associated with each state and observing that excited state contaminations
are strongly suppressed by time slice 21. We then extract the form factors as outlined in
Sec.~\ref{sec:PEVA} for every possible sink time and look for a plateau consistent with a single-state
ansatz.

Inverting through the current requires us to choose our current operators and momentum transfers at inversion time, but
allows us to vary the sink momentum, and thus the source momentum. This gives us access to a range
of values of
\begin{equation}
	Q^2 = \left( E_{\alpha}(\vect{p'}) - E_{\alpha}(\vect{p}) \right)^2 - \vect{q}^2\,,
\end{equation}
in particular, values approaching \(Q^2 = 0\).

\begin{figure}[t]
	\begin{center}
		\includegraphics[width=0.8\textwidth, trim={0 15mm 0 3mm}]{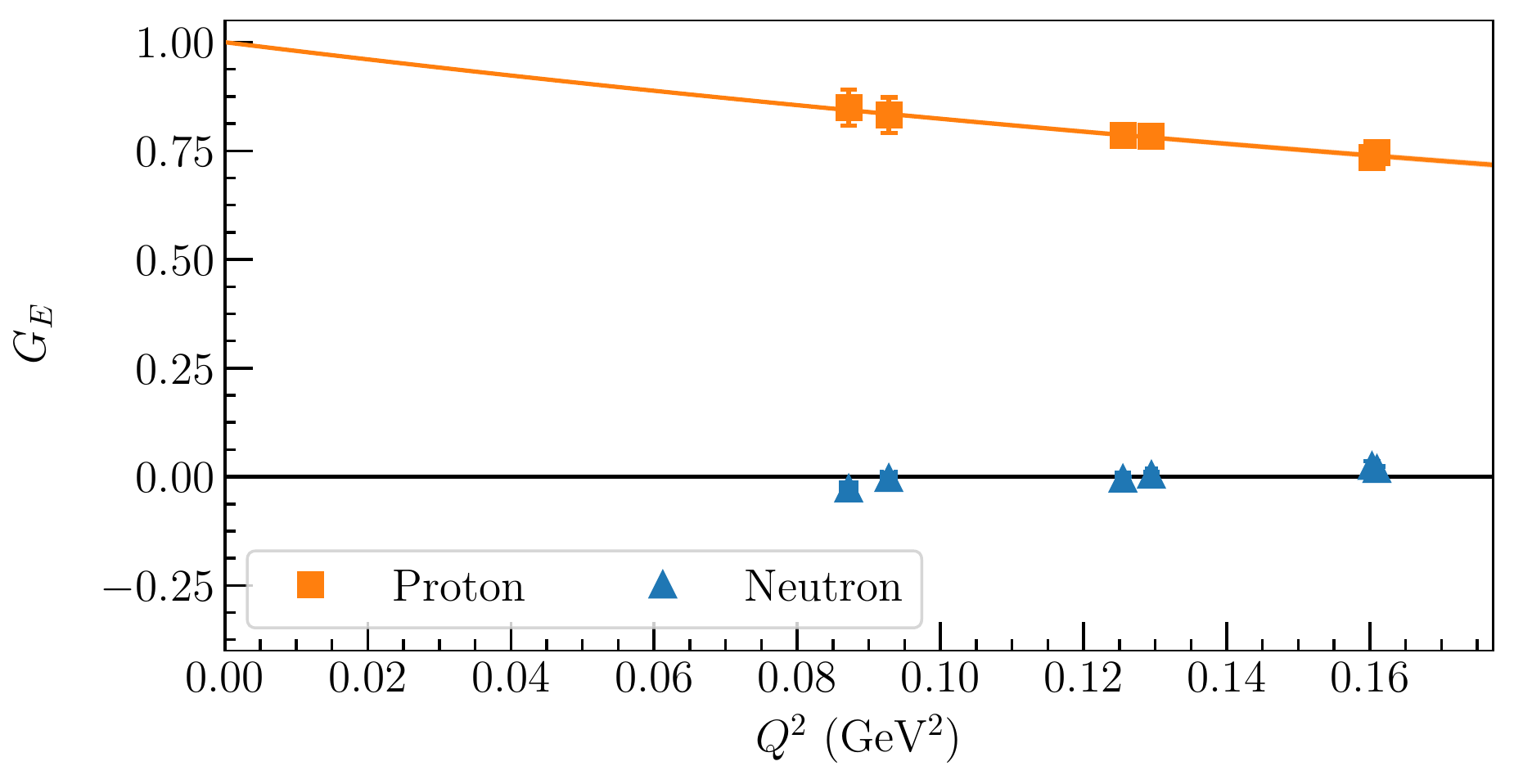}
	\end{center}
	\caption{\label{fig:groundstate:GE}\(G_E(Q^2)\) for the ground state proton and neutron.
	We plot the form factor for the neutron with blue triangles, and the proton with orange squares.
	A dipole fit to the proton's electric form factor is illustrated by the shaded band. This fit
	corresponds to a rms charge radius of \SI{0.689(11)}{\femto\meter}.}
\end{figure}

Beginning with the ground state, in Fig.~\ref{fig:groundstate:GE} we plot \(G_E(Q^2)\) for both the proton
and the neutron with respect to \(Q^2\). Our tactic of projecting different final momenta to access smaller \(Q^2\)
values is seen to work well for the momenta considered. Comparing our extracted form factor for the proton to a dipole ansatz
\cite{Boinepalli:2006xd}
\begin{equation}
	G_{\mathrm{dipole}}(Q^2) = \frac{G(0)}{1+Q^2/\Lambda^2}\,,
\end{equation}
fixing \(G(0)\) to the proton's charge of \(1\), we find good agreement.
The best fit corresponds to a rms charge radius of \SI{0.689(11)}{\femto\meter}, somewhat smaller than
the physical proton radius of \SI{0.8751(61)}{\femto\meter} \cite{Mohr:2015ccw} as expected due to our unphysically
heavy pion mass and well understood finite volume effects \cite{Hall:2013oga}.

\begin{figure}[t]
	\begin{center}
		\includegraphics[width=0.8\textwidth, trim={0 15mm 0 3mm}]{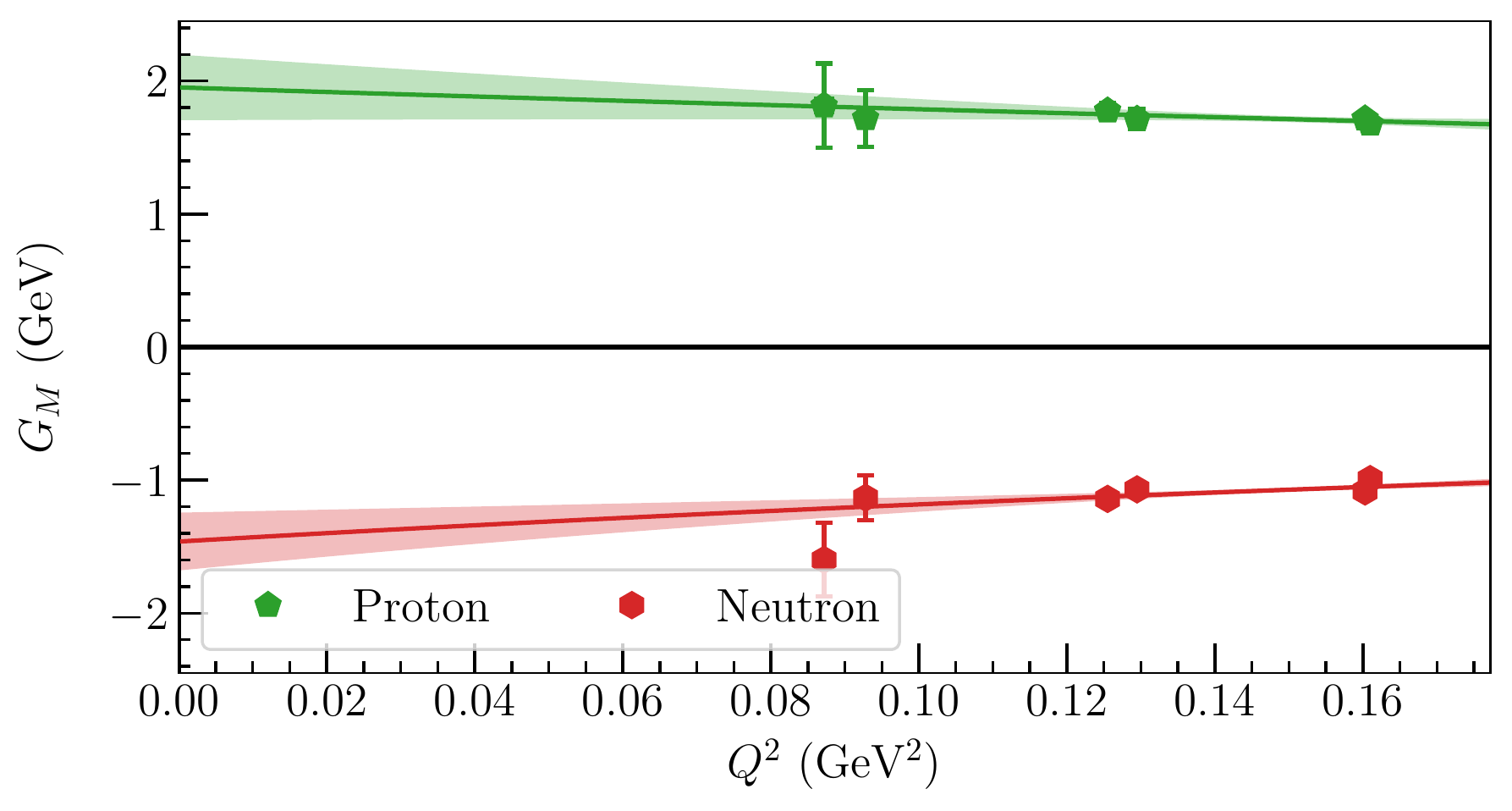}
	\end{center}
	\caption{\label{fig:groundstate:GM}\(G_M(Q^2)\) for the ground state proton and neutron.
	We plot the form factor for the proton with green pentagons, and the neutron with red hexagons.
	Dipole fits to the proton and neutron's magnetic form factors are illustrated by the shaded bands.
	The fit to the proton corresponds to a rms magnetic radius of \SI{0.46(23)}{\femto\meter} and a
	magnetic moment of \(1.95(25) \mu_N\). The fit to the neutron corresponds to a rms magnetic
	radius of \SI{0.72(19)}{\femto\meter} and a magnetic moment of \(-1.46(22) \mu_N\).}
\end{figure}

In Fig.~\ref{fig:groundstate:GM} we plot \(G_M(Q^2)\) of the proton and neutron with respect to \(Q^2\).
Comparing our extracted form factor for the proton to a dipole ansatz, we find good agreement.
The fit corresponds to a rms magnetic radius of \SI{0.46(23)}{\femto\meter} and a magnetic moment of
\(1.95(25) \mu_N\). Comparing our extracted form factor for the neutron to a dipole ansatz, we find a
fit corresponding to a rms magnetic radius of \SI{0.72(19)}{\femto\meter} and a magnetic moment of \(-1.46(22) \mu_N\).

\begin{figure}[t]
	\begin{center}
		\includegraphics[width=0.8\textwidth, trim={0 15mm 0 3mm}]{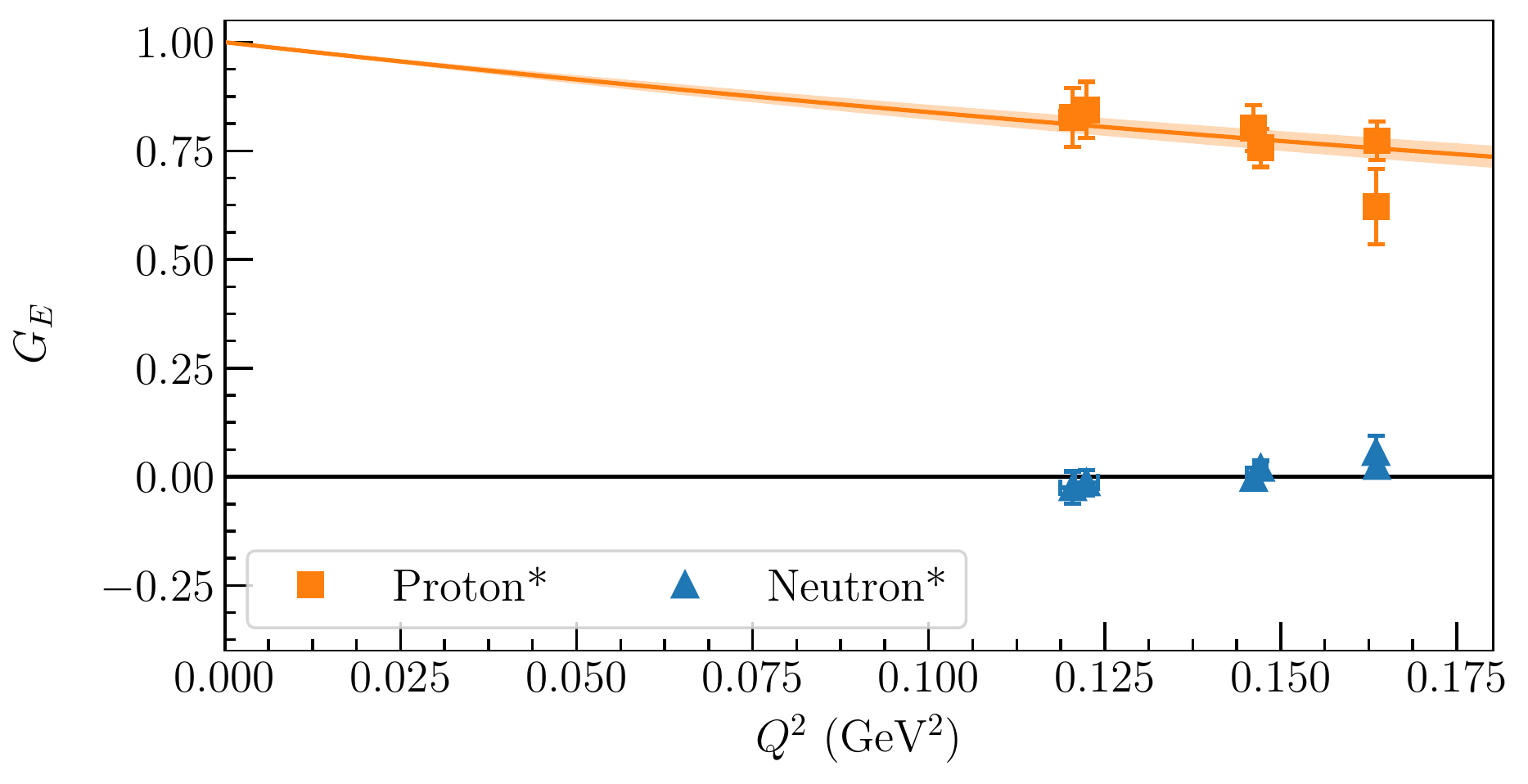}
	\end{center}
	\caption{\label{fig:firstneg:GE}\(G_E(Q^2)\) for the first negative parity excitation of the proton
	and neutron. We plot the form factor for the neutron with blue triangles, and the proton with orange
	squares. A dipole fit to the excited proton's electric form factor is illustrated by the shaded band.
	This fit corresponds to a rms charge radius of \SI{0.655(40)}{\femto\meter}.}
\end{figure}

Moving on to the first negative parity excited state, in Fig.~\ref{fig:firstneg:GE} we plot \(G_E(Q^2)\)
with respect to \(Q^2\). Once again our tactic of accessing multiple \(Q^2\) values through final state momentum
projections is seen to work well. Comparing our extracted form factor for the proton excitation to a dipole ansatz,
we find good agreement with a rms charge
radius of \SI{0.655(40)}{\femto\meter}. The localised nature of this state is evident in its small charge
radius, which is consistent with the radius of the ground state proton.

\begin{figure}[t]
	\begin{center}
		\includegraphics[width=0.8\textwidth, trim={0 15mm 0 3mm}]{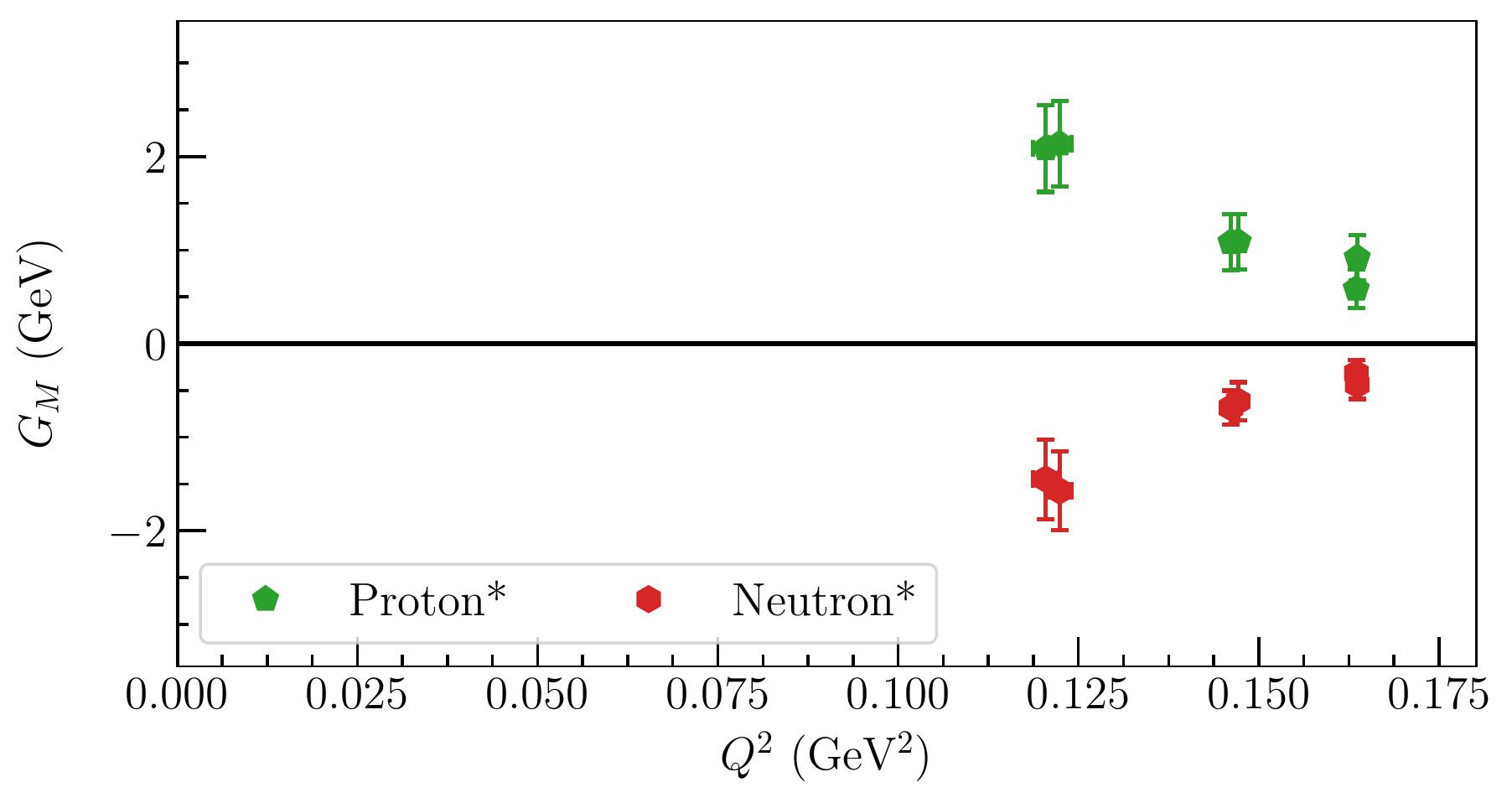}
	\end{center}
	\caption{\label{fig:firstneg:GM}\(G_M(Q^2)\) for the first negative parity excitation of the proton
	and neutron. We plot the form factor for the proton with green pentagons, and the neutron with red
	hexagons. These form factors do not correspond well to any reasonable dipole fit.}
\end{figure}

In Fig.~\ref{fig:firstneg:GM}, we present \(G_M(Q^2)\) for this same excitation. While the largest final state
momenta considered have large errors in this case, we are still able to get some idea of the \(Q^2\) dependence
from the other final state momenta. From these results, it appears that for the first negative parity excitation,
\(G_M(Q^2)\) has a different \(Q^2\) dependence than \(G_E(Q^2)\).

\section{Conclusion}

We have demonstrated the effectiveness of the PEVA technique at extracting baryon form factors. This approach is
effective at removing excited state contaminations of both parities \cite{Menadue:2013kfi, Stokes:2017jiy}.
By varying the sink momentum, we gained access to a range of \(Q^2\) values, allowing us to
compare the \(Q^2\) dependence of the extracted form factors to a simple dipole ansatz, gaining insight into the
structure of the states.

Future work will involve extending this analysis to more pion masses, higher momentum transfers and thus larger
values of \(Q^2\), and other excitations of the nucleon, leading to a comprehensive understanding of excited
nucleon structure. In addition, it will be important to apply these techniques to analysing transition moments
on the lattice, and to analysing non-localised multi-particle channels.

\bibliographystyle{JHEP}
\bibliography{database}

\end{document}

%% file: commands.tex
\usepackage{amsmath}
\usepackage{amssymb}
\usepackage{amsfonts}
\usepackage{amsthm}

\newcommand{\ii}[0]{\mathrm{i}}

\newcommand{\identity}[0]{\mathbb{I}}

\newcommand{\vect}[1]{\mathbf{#1}}
\newcommand{\unitvect}[1]{\hat{\mathbf{#1}}}

\newcommand{\adjoint}[1]{\smash{\overline{#1}}\vphantom{#1}}

\newcommand{\transpose}{^\top}

\DeclareMathOperator{\tr}{tr}

\newcommand{\definedby}{:=\,}

\newtheorem{proposition}{Proposition}[section]